**Epitaxial Graphenes on Silicon Carbide**

Phillip N. First,[1]* Walt A. de Heer,[1] Thomas Seyller,[2] Claire Berger,[3] Joseph A. Stroscio,[4] Jeong-Sun Moon[5]

[1] School of Physics, Georgia Institute of Technology, Atlanta, GA  30332-0430, USA

[2] Lehrstuhl für Technische Physik, Universität Erlangen-Nürnberg, Erwin-Rommel-Str. 1, 91058 Erlangen, Germany

[3] CNRS-Institut Néel, 38042 Grenoble Cedex 9, France

[4] Center for Nanoscale Science and Technology, National Institute of Standards and Technology (NIST), Gaithersburg, MD  20899, USA

[5] HRL Laboratories LLC, Malibu, CA 90265, USA

*E-mail: first@gatech.edu

**Abstract**

The materials science of graphene grown epitaxially on the hexagonal basal planes of SiC crystals is reviewed.  We show that the growth of epitaxial graphene on Si-terminated SiC(0001) is much different than growth on the C-terminated SiC($000\bar{1}$) surface, and discuss the physical structure of these graphenes.  The unique electronic structure and transport properties of each type of epitaxial graphene is described, as well as progress toward the development of epitaxial graphene devices.  This materials system is rich in subtleties, and graphene grown on the two polar faces differs in important ways, but all of the salient features of ideal graphene are found in these epitaxial graphenes, and wafer-scale fabrication of multi-GHz devices already has been achieved.

**Introduction**

The promise of carbon-based nanoelectronics drives a great deal of research on carbon nanotubes, and impressive success has been achieved in creating integrated devices on single nanotubes (see e.g., Avouris et al. and Hersam et al. in this issue). However, the fundamental issue of how to place *billions* of nanotubes without error hinders their acceptance as a technology for large scale integrated electronics, even though the operational characteristics of nanotube devices could exceed the capabilities of silicon. While bottom-up approaches to growing nanotubes in specific configurations continue to be developed, nothing approaching the efficiency of current microelectronic fabrication is on the horizon.

Recognizing this fundamental limitation, Berger et al.[1] chose a different route to carbon nanoelectronics based on lithographic patterning of graphene grown epitaxially on the basal plane of SiC, a wide bandgap semiconductor. To maintain the advantageous properties of nanotubes (e.g., coherent transport,[2] room temperature ballistic transport,[3] size dependent electronic structure[4]) it was proposed to create transistors based on field effect gated graphene nanoribbons.[1] Quantum confinement in such ribbons can create an energy gap that grows with decreasing ribbon width.[5] The gated channels would connect seamlessly to source and drain regions fabricated simply as wider areas of graphene (no confinement gap), thus circumventing the contact issues that also plague nanotubes. This top-down approach to carbon electronics closely follows the present microelectronics paradigm, leveraging continuous improvements in nanolithography and—because the substrate itself is an excellent semiconductor[6]—allowing direct connection to conventional electronics. At the same time, by virtue of the long history of carbon chemistry, the door is open for chemical approaches to patterning, doping, and integration with bottom-up molecular electronics. In this initial work,[1] the experiments showed that good mobility and coherent transport are achievable and demonstrated (if crudely) the essential aspects necessary for large scale integrated graphene nanoelectronics: epitaxial growth on an insulating single-crystal substrate, lithographic patterning, a gate insulator, silicon-scale mobility and field effect gating. Subsequent measurements on improved material showed quantum confinement in a nanoribbon, exceptional carrier mobility, and micrometer-scale coherence lengths.[7]

The key to success of this approach is epitaxial graphene of extraordinary quality. While the growth of "monolayer graphite" has been known in surface science for many years—including growth via thermal decomposition of silicon carbide— the development of epitaxial methods has accelerated since the debut of graphene transport measurements.[1, 8] Improved substrate quality,[9] a better understanding of the growth process,[10] and the discovery of multilayer epitaxial graphene[7, 11] have dramatically expanded the potential of epitaxial graphene on silicon carbide, and multi-GHz devices already have been demonstrated.[12] In this brief review, we discuss the materials science of epitaxial graphene(s) (EG or EGs) on both silicon- and carbon-terminated basal plane surfaces of hexagonal SiC, the status of EG devices, and the potential of EG as a platform for carbon electronics and related technologies. Prior reviews have covered various aspects of this new electronic material in more detail.[13-16] The reader should also be aware that other means of large area graphene growth are under development, most notably chemical vapor deposition on transition metals[17-19] and on copper.[20]

*The Two Faces of SiC*

Silicon carbide has long been of interest as a wide bandgap semiconductor suitable for high temperatures, high electric fields, and high-speed devices.[21] Even for mainstream applications, it is in many ways technically superior to silicon, but presently the device fabrication is more complex.[6] Among its almost 250 crystalline forms, the two of most interest for electronics (and consequently the most available) are the hexagonal 4H and 6H polytypes (energy band gaps of 3.3 eV and 3.0 eV, respectively). Both are formed by stacking basal plane "bilayers" of Si and C, with 0.25 nm *c*-axis spacing and an in-plane lattice constant of 0.307 nm. For the 4H polytype (Figure 1a), the unit cell *c* dimension is 1.00 nm (4 bilayers) and 1.51 nm (6 bilayers) for 6H material (Figure 2a).

The basic mechanism for growing EG on SiC is simply to heat the substrate (in vacuum or inert atmosphere) to temperatures typically in the range 1200 °C to 1800 °C. At these temperatures, Si atoms desorb from the surface (arrows in Figure 1a) and the remaining carbon atoms rearrange to form sheets of graphene. As elaborated in what follows, graphene films are quite different for growth on the silicon terminated $(0001)$ surface (Si-face) versus growth on carbon terminated $\text{SiC}(000\bar{1})$ (C face). For a given temperature and in an open geometry—as would be typical for ultrahigh vacuum (UHV) growth—films grow much faster on the C face. Figure 1b depicts the usual situation (until recently; see below) for each SiC face and shows a feature common to both: Graphene close to the SiC interface is electron doped at typically a few $\times 10^{12}$ cm$^2$V$^{-1}$s$^{-1}$ as a consequence of the work function difference between these materials. The charge-density decay length is approximately one graphene layer,[22, 23] so for multilayer epitaxial graphene (MEG) on C-face SiC, the overlayer planes quickly approach charge neutrality.

**Epitaxial Graphene on SiC$(0001)$**

      Thermal decomposition of SiC to form graphite was first discovered more than a century ago by Acheson,[24] and epitaxial growth of few-layer graphene on SiC$(0001)$ was demonstrated in the 1970s during some of the first studies of SiC surfaces.[25] The basic surface reconstructions and epitaxial layering were illuminated further by scanning tunneling microscopy (STM) and x-ray photoemission spectroscopy (XPS) studies in the 1990s,[26-31] and thermal decomposition of SiC was later proposed as a way to obtain thin graphite films of very high quality, including monolayer graphene.[32, 33] With the advent of lithographic patterning and transport measurements on EG/SiC$(0001)$,[1, 16] the potential to use this material as more than just a smooth substrate was established, and measurements of the valence electronic structure took on a new urgency. Angle resolved photoemission spectroscopy (ARPES) confirmed that the energy vs. momentum relation *E(k)* is linear for monolayer EG[34, 35] and quadratic for the bilayer.[35] The latter work also demonstrated control of the bilayer electronic structure via surface doping and an adsorbate-created electric field, verifying theoretical predictions.[36] These and subsequent measurements established that EG on SiC$(0001)$ has the predicted electronic structure of graphene, although, as

for any complex materials system, there is continuing discussion of data interpretation and structural subtleties.

*Growth and Structure*

The growth of graphene by thermal decomposition is an unusual inverted growth where the graphene thin film forms from C atoms left after desorption of Si from the surface. Since the density of carbon in a graphene sheet is almost the same as the density of C in 3 SiC bilayers, the SiC surface actually recedes as the EG film forms. In conventional vapor deposition, the thin-film quality is controlled largely by a balance between the net deposition flux and the rate of surface diffusion.[37] These parameters are controlled almost independently by adjusting the deposition source and the substrate temperature. However, for EG grown on SiC by heating in UHV, the rate of C "deposition" (i.e., surface C enrichment through loss of Si) and the surface diffusion rate are both determined by the substrate temperature. Surface x-ray reflectivity, STM, and low energy electron microscopy (LEEM), show that such films typically have a relatively high density of SiC steps and pits.[38-42] Even so, good quality EG films have been obtained with average thickness controlled to a fraction of a monolayer. These show the electronic structure characteristic of monolayer graphene,[34, 35, 39, 43] bilayer graphene,[35, 39, 44] and thicker.[39]

Figure 2a shows the basic structure of EG/SiC(0001) after growth, as determined by a number of UHV surface science studies.[22, 31, 45-49] The growth proceeds from step edges, with the decomposing layers first forming a ($6\sqrt{3}$ x $6\sqrt{3}$)R30° carbon rich surface reconstruction known as the "buffer layer" or "layer 0."[32, 45, 46, 48, 50] The exact atomic structure of layer 0 is not known experimentally, but the carbon density is very close to that of a graphene monolayer.[41, 48, 49] Both ARPES measurements[45, 48, 51] and theory[50, 52, 53] suggest that this layer consists of graphene-like $sp^2$-bonded carbon, but the π orbitals interact with the SiC substrate strongly enough to create an energy band gap, determined from experiment to be ≥0.3 eV.[46] The energy gap increases to 1.5 eV with the adsorption of atomic hydrogen, which binds to the buffer layer but less readily to subsequent graphene layers.[54]

With layer 0 providing isolation from dangling bonds of the SiC substrate, layer 1 graphene is the first to display the characteristic graphene honeycomb in STM images. However, the imaging is very dependent on the sample-tip bias voltage, with the SiC interface states dominating images beyond ±0.5 V and pristine graphene imaged at low bias.[22, 46, 55] Figure 2b shows an image acquired at the transition between these imaging conditions. As a result of structure in the underlying buffer layer reconstruction, the surface of layer 1 appears corrugated[22, 46, 53, 55] (at low tunnel voltages) with a period of 1.85 nm and peak-to-valley amplitude of typically 40 pm to 60 pm. X-ray reflectivity measurements[56] indicate that this is a real geometric distortion. Interface states are largely absent in images of layer 2, although the corrugation is still apparent, with reduced amplitude.[22, 57] It is typical to image only ½ of the atoms in this layer ("3-for-6" imaging) due to the ordered stacking of layer 2 on layer 1, but this also depends on the tunnel bias,[58] and transitions from 3-for-6 to 6-for-6 imaging are occasionally observed. These may indicate stacking transitions.[22, 59] A final important observation is that graphene grows continuously over steps in the substrate.[1, 22, 46, 60, 61]

It is possible to decouple surface diffusion from the rate of carbon enrichment by controlling the net flux of Si atoms leaving the surface. This can be accomplished in different ways, e.g., by creating a closed and Si-rich environment,[16] by directly controlling the Si vapor pressure,[62] or by using a buffer gas to increase the probability of desorbed Si atoms returning to the surface.[10, 63] Figure 2c shows the result of graphenizing SiC(0001) in an atmosphere of 100 kPa argon. The surface shows substantial bunching of SiC steps, with extended flat terraces covered by layer-1 material, as shown by the height profile in Figure 2d. The terrace lengths are hundreds of micrometers, while the ≈2μm mean terrace width is determined by the miscut angle of the substrate. Bilayer graphene is found only at the step edges, suggesting that the homogeneity can be further improved by tight control of the step density.[63] This was recently confirmed via growth on substrates with various off-axis angles.[64] Clearly, wafer scale graphene is achieved via epitaxial growth on SiC(0001).

*Electronic Properties and Transport*

As indicated above, the electronic properties of EG on SiC(0001) are layer dependent. An important fact is that the buffer layer has an energy gap at $E_F$, so transport experiments and valence spectroscopies measure the effect of the graphene layers. Figure 3a shows ARPES data from layer-1 EG on SiC(0001). The experimental $E(k)$ is linear, with a characteristic band velocity consistent with the band structure of an ideal monolayer. Close examination of the spectrum reveals a small shift of the energy bands above the Dirac (charge neutrality) point $E_D$ relative to the bands below $E_D$. This has been ascribed to many-body interactions[43, 65] or to the creation of a small band gap.[51, 66] Resolution of this issue remains a focus of experiment[67-69] and theory.[52, 53, 70, 71]

The parabolic energy bands of layer 2 graphene are apparent in Figure 3b, as is the lower energy split-off band. These observations are as predicted for bilayer graphene.[36, 72] The small energy gap centered around -350 meV is due to the interface electric field shown schematically in Figure 1b; it can be driven to zero by balancing the interface field with an electric field contributed by surface adsorbates.[35] Carrier density is also a layer dependent quantity in EG.[39] The tunneling spectra in Figure 3c show how the charge neutrality point shifts with respect to the Fermi level (zero tunnel bias) for successive EG layers on SiC(0001).[22] The corresponding decay length for the charge density [proportional to $(E_F-E_D)^2$] is somewhat larger than 1 EG layer.

EG is well suited to macroscopic probes of electronic structure. However, unlike micro-cleaved graphene flakes on $SiO_2$/Si, EG has no built-in backgate to enable continuous adjustment of the carrier density. A backgate is typically not necessary for devices, but clearly would be convenient for more complete studies of transport properties in this unique 2D system. Nevertheless, even the first magnetotransport studies of EG[1] showed that the carrier mobility is large (1100 $cm^2V^{-1}s^{-1}$ at 4 K, $n$=3.6x10$^{12}$ cm$^{-2}$; note that for graphene, mobility increases as the carrier density $n$ decreases) and that the system has a high degree of coherence. Shubnikov-de Haas oscillations (i.e., resistance oscillations with magnetic field due to quantization of cyclotron orbits) also were observed[1] and later shown to imply a Berry phase of $\pi$,[16] characteristic of monolayer graphene. More recently, monolayer sample mobilities over 2000 cm$^2$V$^{-1}$s$^{-1}$ (27 K; 900 cm$^2$V$^{-1}$s$^{-1}$ at 300 K) have been achieved[10] for high electron densities $n \approx 1 \times 10^{13}$ cm$^{-2}$ and almost 30 000 cm$^2$V$^{-1}$s$^{-1}$ for an electron density of $n = 5.4 \times 10^{10}$ cm$^{-2}$, reduced through adsorption of an acceptor molecule.[73] Substrate steps have been found to have little effect on the mobility,[73] but may affect the level of self doping.[74]

Shubnikov-de Haas oscillations (SdHOs) have been followed to very high magnetic fields[73, 75, 76] where quantized transverse (Hall) resistance is also found (acceptor-doped samples reduce the field scales to below 8 T.[73]) The phase of the SdHOs shows that this half-integer quantum Hall effect (QHE) is the same as measured earlier in graphene flakes.[77, 78]

**Epitaxial Graphene on SiC$(000\bar{1})$**

Graphene also grows on the carbon terminated $(000\bar{1})$ surface (C face) of silicon carbide. As for the Si face, growth progresses by thermal decomposition in vacuum and or in an inert gas environment. However, since the first observations,[25] it has been recognized that graphene grows quite differently on the two different surfaces, with Si-face material clearly epitaxial [e.g., showing sharp spots in low-energy electron diffraction (LEED)] while UHV-grown C-face graphene shows many rotational domains, or even sprouts nanotubes.[79] Control of the C-face graphenization can be achieved by enclosing the SiC substrate in a furnace.[16] This method produces high quality multilayer epitaxial graphene (MEG) with unique layer-stacking that results in *n*-layer MEG behaving effectively as *n* independent graphene monolayers.[11] However, as indicated in Figure 1b, those layers that lie close to the SiC interface are highly electron-doped, with the charge density decay length approximately one layer, similar to the Si-face material.[23] Consequently, a single layer has the highest carrier density and the highest conductivity. This "transport layer" dominates conventional magnetotransport measurements, whereas most electron and optical spectroscopies measure the nearly neutral overlayers. The effect of the overlayers on the magnetotransport is subtle (see below); a significant advance in the growth of graphene on C-face SiC has been the recent achievement of true monolayer graphene.[80]

*Growth and Structure*

Multilayer growth on the C face has been the norm until lately, so detailed studies of the graphene/SiC($000\bar{1}$) interfacial atomic structure have been limited to either UHV-grown samples or surface x-ray scattering.[14, 38, 56] UHV studies show that at low temperatures (≈1100 °C) the clean ($000\bar{1}$) surface has a 3x3 reconstruction that coexists with a 2x2 reconstruction once the surface has been graphenized with a single graphene overlayer.[81] The STM data[81] suggest a weak coupling of the first graphene layer to the reconstructed substrate in agreement with photoemission work[48] and DFT calculations,[82] which also indicate linear π-band dispersion at the K point. For furnace grown MEG, the in-plane atomic arrangement is unknown, but x-ray reflectivity suggests that the first graphene layer binds tightly to the topmost SiC bilayer, which itself may be carbon rich.[56] This configuration may prove to be essential for isolating subsequent layers from interaction with the substrate. Clearly, future research will need to reconcile these different findings for graphene grown on the C face by different methods.

Beyond the initial graphene formation, UHV-grown material on C-face SiC shows little orientational order, and tends to form 3D structures.[79] In contrast, for furnace-grown C-face graphene, successive layers maintain their planarity, and the registry of adjacent graphene layers is dramatically different than for EG grown on the Si face. Whereas Si-face graphene exhibits the Bernal (ABAB…) stacking of graphite, layer stacking on the C face is complex. LEED, x-ray scattering, and STM show that adjacent layers in C-face epitaxial graphene are typically rotated with respect to one another at angles not associated with Bernal stacking (i.e., the relative angles are not 60°). X-ray diffraction and LEED indicate that the preferred rotation angles lie near 0°—in a band of ±5°, as shown in the inset to Figure 4b—and at 30° with respect to the ⟨$21\bar{3}0$⟩ direction of SiC.

Integrated diffraction intensities show that the ≈0° and 30° orientations occur with equal probability and that the rotated layers are interleaved, as opposed to forming distinct domains.[11, 14] In other words, most graphene sheets register at an angle of ≈30° relative to adjacent layers. Figure 4a shows the most frequent layer alignment, which may be favored due to an epitaxial match with the SiC substrate, where the layers form.[11] Angles far from these values are detected only infrequently in high-temperature grown material,[56] but are commonly found in UHV where the growth temperature is lower.[83]

The rotational stacking gives characteristic moiré images in STM (Figure 4b,c) where the contrast in apparent height is caused by periodic differences in the local stacking structure of the top few graphene layers. Double moiré patterns—involving at least 3 graphene layers—also are observed.[84] ARPES studies find minimal occurrence of Bernal-stacked layers in the multilayer film.[85] This indicates that the rotated graphene layers in high temperature furnace-grown MEG are *not* distributed randomly in an otherwise graphitic film. Based on a measured rotational fault density of one every 2.5 graphene layers,[56] a random fault model would predict a Bernal stacking fraction near 50 %, which is far larger than measured in ARPES.[85] It remains for future experiments to determine the detailed sequence of layer rotations, which may be tied to the kinetics of graphene growth at the SiC interface.

Finally, we note that the MEG layers are found to be extremely flat, and continuous over substrate steps and rotational domain boundaries.[56, 84] As a result of the thermal expansion mismatch between SiC and graphene, isolated nanometer-high folds of the graphene occur every 10 μm to 20 μm in furnace-grown material, but the graphene remains continuous through these features (see, e.g., Figure 7a).[14, 86, 87] Thus the topmost layer of MEG (at the least) is continuous over the entire surface of the SiC crystal.

*Electronic Properties and Transport*

The unusual rotational stacking has important consequences for the electronic properties of graphene multilayers grown on $SiC(000\bar{1})$. As shown theoretically and experimentally,[11, 85, 88, 89] the electronic structure of two incommensurately stacked graphene layers is essentially equivalent to that of two freestanding graphene monolayers, with a well defined Dirac cone [linear $E(k)$] near the charge neutrality point.[85]. Hence the material is appropriately called multilayer epitaxial graphene, and not graphite. The unperturbed Dirac cone results in the same single-Lorentzian G′ (or 2D) peak in the Raman spectrum of MEG[90] as observed previously in single-layer graphene flakes.[91]

Angle resolved photoemission spectroscopy of the topmost (neutral) layers in the MEG stack (Figure 4d,e) reveals directly the decoupled nature of the layers.[85] In contrast to ARPES of the Bernal bilayer on Si-face EG (Figure 3b), the Dirac cones of the MEG layers remain unperturbed and distinct from one another. The $k_\perp$ displacement of the cone sections in Figure 4d,e is due to the rotation angle between layers.

Methods based on the quantization of cyclotron orbits in a magnetic field have long been used to obtain very precise characterization of the electronic structure of materials and two-dimensional electron- or hole-gas systems (2DEGs). In normal 2DEGs the dispersion is parabolic (i.e., the carriers have finite effective mass *m*), giving a constant density of states versus energy. In SI units, the cyclotron orbit frequency in a magnetic field *B* is $\omega = \frac{eB}{m}$, where *e* is the carrier charge and *m* the carrier effective mass. This gives rise to a density of states consisting of discrete "Landau levels" (LLs) each of identical degeneracy and equally spaced in energy: $E_n = \left(n + \frac{1}{2}\right)\hbar\omega$, with *n* the integer quantum number of the LL. In ideal graphene, the density of states increases linearly with energy, leading to a qualitatively different Landau-level spectrum:

$$E_n = \pm c^* \sqrt{2e\hbar B |n|}; \qquad (1)$$

where $c^*$ is the characteristic band velocity of graphene. Not only are the LLs unequally spaced, but the energy of the *n=0* level does not depend on the magnetic field. This essential feature of the graphene LL spectrum is due to the nontrivial Berry's phase[92] and revealed in the QHE.[77, 78]

Infrared spectroscopy of MEG in a magnetic field measures transitions between LLs and precisely confirms the $\sqrt{B}$ dependence in Equation 1, consistent with the electronic decoupling of layers[93, 94] and different than 3D graphite, even at a thickness of 100 layers.[95] Even more striking, these experiments show that the Landau levels can be resolved in relatively weak magnetic fields, all the way to room temperature.[94] The minimum field for observable transitions implies a carrier density of ≈$5 \times 10^9$ cm$^{-2}$ in the overlayers and mobility greater than 250 000 cm$^2$V$^{-1}$s$^{-1}$. This would be maintained to room temperature based on the measured small and almost temperature independent electron-phonon coupling.[94]

The spectrum of LLs in the top layer of MEG has been measured directly using low temperature scanning tunneling spectroscopy (STS).[84] Figure 5a displays a cartoon of the orbit quantization condition and a Landau-level wave function overlaid on an actual STM topograph of the region studied. Both the graphene atomic structure and the moiré modulation of the apparent height are visible in the topograph. STS results are given in the remaining figure panels. The inset to Figure 5b shows schematically the discrete LL states in momentum space, while the data in that panel are "tunneling magnetoconductance oscillations" (TMCOs) measured by ramping the magnetic field with the STM sample-tip bias held fixed. Maxima of the TMCOs occur when successive Landau levels coincide with the tunnel bias energy. This is the same mechanism underlying Shubnikov-de Haas oscillations, but unlike SdHOs, the energy probed by TMCOs is not restricted to the Fermi energy; they can be used to map the energy bands directly, as shown in Figure 5c (note that both filled states and empty states are probed by STS).

The tunneling conductance versus voltage spectrum (*dI/dV* vs. *V*) is directly comparable to the local density of states versus energy. Thus, in the *dI/dV* spectrum of Figure 5d, the LLs at *B=5T* appear as sharp peaks, with essentially zero density of states between neighboring LLs, until the Lorentzian tails of the peaks begin to overlap. Lifetimes derived from the Landau-level widths (0.4 ps for the *n=0* Landau level)[84] compare very favorably with carrier lifetimes in high-mobility samples of suspended graphene.[96] A small peak splitting of the *n=0* LL requires more investigation, but may be due to electron-electron interactions.

As indicated above, conventional transport experiments are dominated by a highly doped layer near the SiC interface (SdHOs determine the typical carrier density to be *n*-type, $10^{12}$ cm$^{-2}$ to $10^{13}$ cm$^{-2}$). C-face samples that consist of *only* this single layer of graphene (Figure 6a) display well-developed plateaus in the Hall resistance $\rho_{xy}$ as successive LLs are filled with decreasing *B*.[80] Corresponding SdHOs in the magnetoresistance $\rho_{xx}$ determine a Berry phase of π, and reach zero resistance for the *n=1* and *n=0* Hall plateaus (QHE filling factors $\nu = 6$ and $\nu = 2$). These features are characteristic of the half-integer quantum Hall effect observed previously in single-layer graphene flakes on SiO$_2$/Si substrates.[77, 78] It is noteworthy that the QHE is beautifully demonstrated in Figure 6a, even though the graphene monolayer spans several steps in the substrate and the processing contamination is relatively high. The measured mobility of this sample is 20 000 cm$^2$V$^{-1}$s$^{-1}$ at T=4 K and 14 000 cm$^2$V$^{-1}$s$^{-1}$ at T=300K.

Magnetotransport measurements on two-dimensional MEG samples are enigmatic: $\rho_{xy}(B)$ is essentially featureless and $\rho_{xx}(B)$ shows very weak SdHOs (Figure 6b) that don't develop into the QHE, even though the transport mobilities are high.[16] Nevertheless, the measured Berry phase of π shows that the transport layer also has the electronic characteristics of single-layer graphene due to the layer decoupling. Quenching of the QHE in MEG samples has been explained as a consequence of field-dependent scattering into the *n=0* LL of the undoped overlayers, which is always coincident with the Fermi energy in the transport layer.[97] On the other hand, transport measurements of relatively narrow MEG ribbons (Figure 6c) show well developed SdHOs.[7] Structure in $\rho_{xy}(B)$ is not fully understood, but features that may be related to quantum Hall plateaus are observed (Figure 6c).

As anticipated,[1] MEG ribbons show temperature-dependent electronic confinement for widths under a few hundred nanometers.[7] Significant interference effects are found, resulting from micrometer-long phase coherence lengths. Weak

*anti*localization was also predicted for graphene[98] due to the suppression of backscattering.[92] This was first observed in wide MEG ribbons.[99]

All of the remarkable properties of graphene have been demonstrated for epitaxial graphene grown on $SiC(000\bar{1})$. This material will continue to be useful for advancing the science of graphene, and we anticipate that both single-layer and multilayer graphene on the carbon-terminated face will find many applications in electronic and electromagnetic devices.

**Devices**

Perhaps the ultimate potential for graphene devices and sensors lies in completely new concepts that will exploit the unique properties of this novel material. However, a device platform that closely follows the present electronics paradigm is beneficial for rapid acceptance and further development, especially if it scales to nanometer size more effectively than silicon. Epitaxial graphene on SiC has many attractive properties for conventional high-speed and nanometer-scale field-effect transistors: High carrier mobility, ballistic and coherent conduction, small temperature coefficient of resistance, high maximum current density, chemical inertness, size-tunable electronic structure, and direct growth on a single-crystal semiconductor—obviating the need to transfer a wafer-size atomic monolayer to another substrate. Large-area patterning can be performed using the established methods of microelectronics. If required, connection to conventional electronics also could be accomplished in various ways, such as through SiC devices or III-nitride devices, for which SiC is also an excellent substrate.

The various stages of processing to produce wafer-scale epitaxial graphene transistor arrays are shown in Figure 7: Growth of a uniform graphene sheet (multilayer or single-layer) on silicon carbide (Figure 7a); lithographic patterning (Figure 7b); dielectric deposition (e.g., hafnia or alumina; not shown); and finally applying the leads (Figure 7c). The magnetotransport characteristics of a top gated Si-face epitaxial graphene field-effect transistor (EGFET) are plotted in Figure 7d. The figure shows the characteristic polarity effect: for negative gate voltages the carriers are holes while for positive gate voltages they are electrons (evident from the Hall effect). A maximum in the channel resistance occurs near the Dirac point. This wide-channel transistor has an on-to-off ratio of about 30.

Though straightforward, all of the processing steps are challenging and affect the graphene mobility. Furthermore, a number of auxiliary materials issues need to be investigated more thoroughly, such as low-resistance contacts, low loss nonhysteretic dielectrics, and perhaps a native dielectric.[100] Note that global backgating, as employed so successfully in studies of 2DEG physics, is less useful for large scale graphene-based electronics.

Large arrays of EGFETs have been produced on both Si- and C-face SiC using the processing steps given above.[101] Although the transistors were rudimentary, they did provide proof of principle for large-scale device manufacturing. In these devices the graphene transistor channels were too wide (10 µm) to exhibit the quantum confinement bandgap, so that the off-to-on resistance ratios were unimpressive (≈10). Confinement effects will be enhanced for narrow channel EGFETs (≤ 10 nm), as already demonstrated in exfoliated graphene transistors. An all graphene transistor with graphene side gates also has been demonstrated.[102]

While the low on/off ratio of the first wide-channel EGFETs is problematic for logic devices, there is an entire class of high frequency analog transistors that require only a net current gain and not a large on/off ratio. For these devices, an on/off ratio of ≈20—as observed in single-layer EGFETs[102] — already suffices, but the carrier mobility must be high to achieve high frequency operation. Graphene exceeds the highest carrier mobility (electron or hole) of any semiconductor: Over $10^5$ cm$^2$V$^{-1}$s$^{-1}$ at room temperature for suspended graphene[96] or MEG.[94] This is about 10 times greater than that of state-of-the-art high electron mobility transistors (HEMTs) made from lattice-matched InP,[103] the current material of choice for low-noise amplifiers in millimeter wave (mmW) and sub-mmW receiver applications. The saturation velocity of graphene is also estimated to be 3 to 5 times larger than that of lattice-matched InP HEMTs,[104] making it an attractive candidate for mmW and sub-mmW operation (ballistic transport would push operating frequencies still higher). For low-noise receiver applications, a combination of high transconductance and low access resistance relative to the input gate capacitance could provide an excellent noise figure at mmW frequencies such as W band (75 GHz to 100 GHz) and beyond.

In fact it is likely that high frequency transistors will become the first application of graphene based electronics. These devices pose additional technological challenges, but lately their development has seen rapid progress, with transistors already operating over 10 GHz, as shown in Figure 8.[12] Table I gives a comparison of the speed metric $f_T \cdot L_g$ among different transistor technologies ($f_T$ is the unity-gain frequency, $L_g$ the gate length). Even in its very early development, EG on SiC is a competitive technology, and the $f_T \cdot L_g$ product of RF-EGFETs is expected to improve substantially as the quality of the EG layer and transistor fabrication improve, reducing the parasitic charging delay.

SiC is an excellent low-loss substrate for these high-frequency devices because its optical phonon energy is high (115 meV to 120 meV, 2x larger than $SiO_2$). Scattering from substrate optical phonons can limit the mobility of graphene carriers, especially at high temperatures. Effective passivation of the EG/SiC interface may further reduce the effect of the substrate,[105] and it is even feasible to remove the substrate completely, as demonstrated recently by the creation of freestanding epitaxial graphene membranes.[106]

**Conclusion**

Graphene grown epitaxially on silicon carbide displays the predicted properties of ideal graphene and attains carrier mobilities equivalent to suspended graphene. A great deal of materials development remains to be done, especially for nanometer-scale logic devices, but high frequency transistors already have been fabricated on the wafer scale. Epitaxial graphenes on silicon carbide comprise a versatile materials system that will deliver on the promise of graphene-based electronics.

**Acknowledgments**

We thank E. Conrad and A. Zangwill for helpful discussions. PNF acknowledges support from NSF [DMR-0804908; DMR-0820382 (MRSEC)] and the Semiconductor Research Corporation (NRI-INDEX). WdH and CB were supported by NSF (DMR-0820382), NRI-INDEX, and the W. M. Keck Foundation. TS acknowledges support by the Deutsche Forschungsgemeinschaft (DFG), by the Bavaria California Technology Center (BaCaTeC), and by the excellence cluster "Engineering of Advanced Materials" (EAM) at the Friedrich-Alexander-University Erlangen-Nürnberg. Work at HRL (JSM) was supported by DARPA contract number N66001-08-C-2048, monitored by Dr. Michael Fritze.

**Figure Captions**

Figure 1. Schematic structure of silicon carbide and the growth of epitaxial graphene. (a) 4H-SiC. Yellow and green spheres represent Si and C atoms, respectively. At elevated temperatures, Si atoms evaporate (arrows), leaving a carbon-rich surface that forms graphene sheets. (b) At a typical growth temperature, few graphene layers are formed on the Si-terminated face and substantially more on the C-terminated face. The graphene layer(s) close to the SiC interface is electron doped, while the overlayers are essentially undoped (the measured charge decay length is approximately one layer [22, 23]).

Figure 2. Epitaxial graphene on 6H-SiC(0001) (Si-terminated face). (a) Schematic of EG layer structure. (b) Left to right: Scanning tunneling micrographs of the carbon-rich buffer (layer 0); the first layer with essentially graphene electronic structure (layer 1; imaged at 0.4 V, 100 pA—tunnel conditions sensitive to both graphene and to subsurface interface states); the second graphene layer (layer 2), which shows 3-for-6 imaging and a small variation of the atomic heights due to the 1.8 nm SiC "6x6" corrugation. (c) Left: AFM image of 1.2 monolayers EG grown on 6H-SiC(0001) under 90 kPa Ar pressure. Right: Surface height profile along line AB in the image. The profile shows bunching of SiC bilayer steps to form larger steps ≈54 bilayers high (nine 6H unit cells) with terraces covered by monolayer EG. Integers at top label the number of graphene layers. (d) LEEM image of the same EG sample revealing monolayer coverage on the terraces and bilayer/trilayer growth at the step edges. Part (a) adapted from [22], (b) adapted from [107], and (c), (d) from [10].

Figure 3. Layer-dependent electronic structure of epitaxial graphene on SiC(0001). (a) ARPES of a sample dominated by layer 1. (b) ARPES of a sample dominated by layer 2. (c) Scanning tunneling spectroscopy showing the shift of the Dirac point (arrows) relative to the Fermi energy (zero sample bias) for graphene layers 1 through 4. The shifts imply that the charge density decreases in successive layers. Parts (a) and (b) adapted from Ref. [35]; (c) from Ref. [22].

Figure 4. Rotational stacking faults in multilayer epitaxial graphene (MEG) on SiC(000$\bar{1}$). (a) *($\sqrt{13}$ x $\sqrt{13}$)R46.1°* unit cell alignment of two graphene sheets. This particular moiré cell is also commensurate with the SiC substrate, which may account for its prevalence in the overlayer stack. (b) STM topograph showing the moiré superlattice on the top layer of a nominally 10-layer MEG sample. *Inset:* X-ray diffraction intensity (azimuthal scan) from graphene overlayers aligned near the 0° (SiC) azimuth. (c) High resolution image of the *($\sqrt{13}$ x $\sqrt{13}$)R46.1°* superlattice. The atomic height corrugation (15 pm to 20 pm peak-to-valley in the raw data) has been reduced by Gaussian smoothing in order to make the longer-period moiré pattern (≈8 pm peak-to-valley in the raw data) more visible. (d) ARPES energy bands of an 11-layer MEG film measured at a temperature of 6 K. The wave vector scan is perpendicular to the SiC ⟨10$\bar{1}$0⟩ direction through the *K* point. Three linear Dirac cones (one faint) can be seen. (e) A momentum

distribution curve (MDC) at binding energy ($E_F - 0.675$ eV) shows intensity due to all three cones. Heavy solid line is a fit to the sum of six Lorentzians (thin red lines). (a) through (c) reprinted from [11]; (d) and (e) reprinted from [85].

Figure 5. Electronic structure of MEG (top layer) from scanning tunneling spectroscopy (STS) performed in a magnetic field. (a) Foreground shows a cartoon of the quantized cyclotron orbits (Landau levels) probed by STS. In the background is an STM topograph of the sample showing the graphene atomic honeycomb and a small (≈0.01 nm) modulation in the apparent height due to the moiré alignment of layers. (b) Inset: Landau level (LL) energy structure. Landau levels lie at discrete energies determined by continuity of the carrier wavefunction around a cyclotron orbit. The data shows tunneling magnetoconductance oscillations (TMCOs) detected in the tunneling dI/dV. Changing the magnetic field *B* expands the fan of LLs, resulting in a peak in the dI/dV when a LL sweeps through the energy $eV_B$ set by the fixed sample-tip bias $V_B$. (c) At a fixed magnetic field, the LLs appear as peaks in the dI/dV as the sample bias is changed (*B*=5 T for this spectrum). The inset shows that the LL energies are fit by a model of single-layer graphene ($\gamma_1=0$). (d) Both the TMCO measurements of (b) and the conventional STS in (c) imply a linear *E(k)* relation. Shown here are the TMCO energy bands. Part (a) courtesy K. Kubista; (b) through (d) reprinted from [84].

Figure 6. (a) Quantum Hall effect in single-layer epitaxial graphene measured at 1.4 K. Red line shows the Hall resistance with characteristic Hall plateaus at $\rho_{xy} = (h/4e^2)/(n + \frac{1}{2})$ where *n* is the Landau level index. Black line shows oscillations in the magnetoresistivity $\rho_{xx}$ and zero resistance for *n*=0 and *n*=1 Landau indexes. *Inset:* AFM image of the Hall bar (scale bar = 2 μm) patterned over several SiC steps. White specks are electron-beam resist residue; white lines are pleats in the graphene. (b) High field magnetoresistance variation for a 5 μm wide ribbon after subtracting a smooth background. Temperatures of 4 K, 10 K, 20 K, 30 K, 50 K, and 70 K show SdHOs of progressively decreasing amplitude. (c) Magnetotransport of a 6μm x 0.5μm MEG Hall bar measured at temperatures of 4 K, 6 K, 9 K, 15 K, 35 K, and 58 K. Components of the resistivity tensor are shown. Part (a) adapted from [80], part (b) from [16], and part (c) adapted from [7].

Figure 7. (a) AFM image of furnace-grown C-face MEG. The surface is flat except where MEG drapes over steps of the substrate and over folds or pleats (white lines) that form to relieve stress accrued between MEG and the SiC substrate as they cool. (b) SEM picture of a patterned Hall-bar structure. The ribbon is patterned on a single terrace, with graphene pads extending towards the Pd/Au contacts. (c) Example of integrated structures on a SiC chip, featuring a pattern of 100 ribbons. The background contrast is an artifact from the tape on the back of the transparent SiC chip. (d) Resistivity $\rho_{xx}$ and Hall resistance $\rho_{xy}$ as a function of gate voltage at 5 T and 300K for a 3.5 μm x 12.5μm graphene Hall bar on Si-face SiC. The resistivity peaks when $\rho_{xy}$ changes sign. *Inset:* optical image of the gated structure. Three gates ($G_1$, $G_2$, $G_3$) deposited on top of the dielectric (light brown rectangle) partially cover the ribbon that lies between current leads *I*, and voltage probes *V*. Adapted from [102], "Top and side gated epitaxial graphene field effect transistors," Copyright Wiley-VCH Verlag GmbH & Co. KGaA. Reproduced with permission.

Figure 8. (a) 50mm graphene wafer processed by standard lithographic techniques. (b) Scanning electron micrograph of a 2 μm x 12 μm graphene FET. (c) Measured |H$_{21}$| and unilateral gain (*U*) as functions of frequency for 2 μm x 12 μm graphene FETs measured at $V_{ds}$ = 5 V and $V_{gs}$ = -2.5 V. An extrinsic cutoff frequency $f_T$=4.1 GHz is extracted, yielding an extrinsic $f_T \cdot L_g$ of 8.2 GHz·μm. The extrinsic $g_m$ is 195 mS/mm. A maximum oscillation frequency of $f_{max}$ =11.5 GHz is extracted from the unilateral gain (*U*) with a slope of -20 dB/decade. (d) *n*-FET and *p*-FET device characteristics. Adapted from [12] and HRL press releases.

Table I. Speed Comparison of Semiconductor Transistor Technologies.

| Technology | $f_T \cdot L_g$ (GHz·μm) |
|---|---|
| InP | 22 |
| ITRS Bulk NMOS | 9 |
| SOI (90 nm) | 11 |
| HRL Graphene (2008) | 10 |

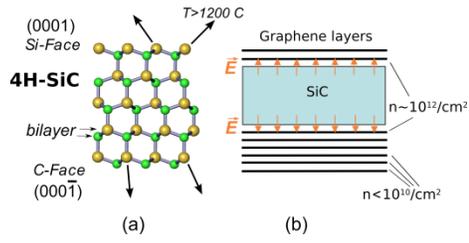

**Figure 1**

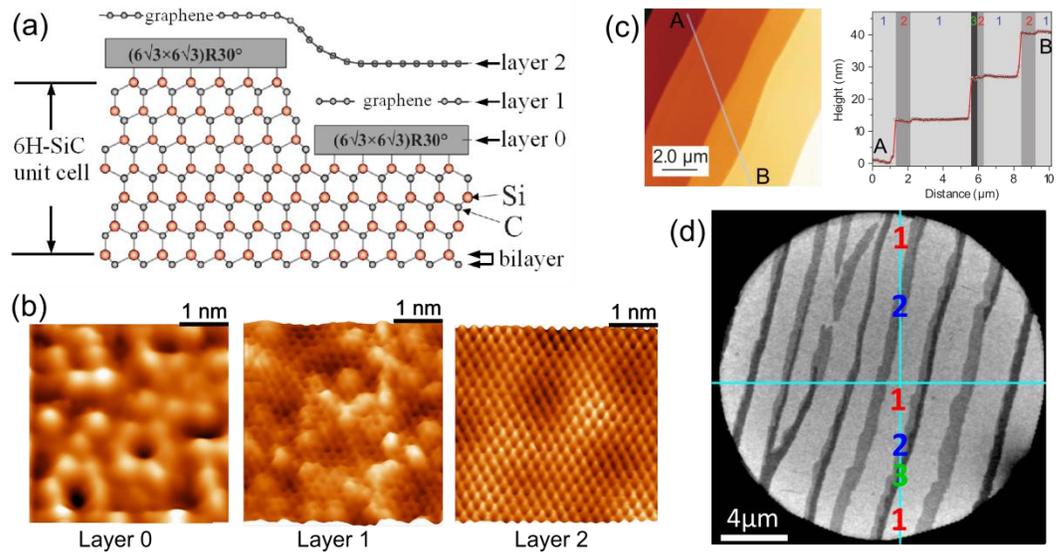

**Figure 2**

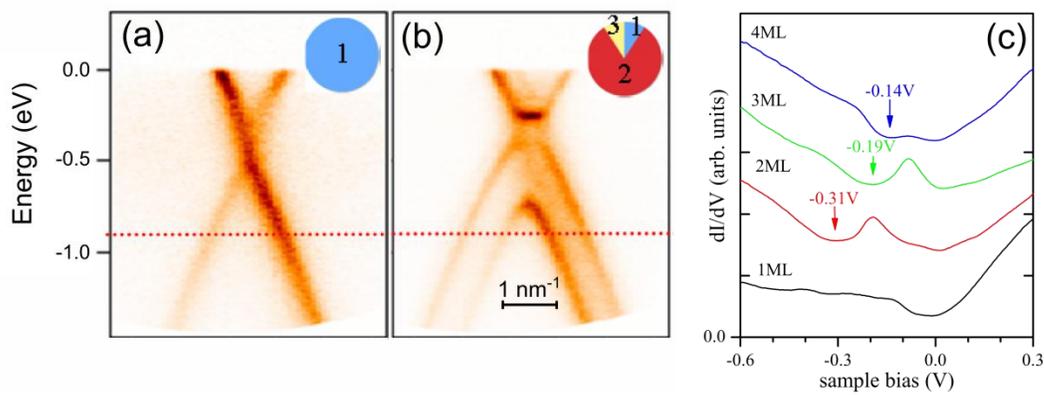

**Figure 3**

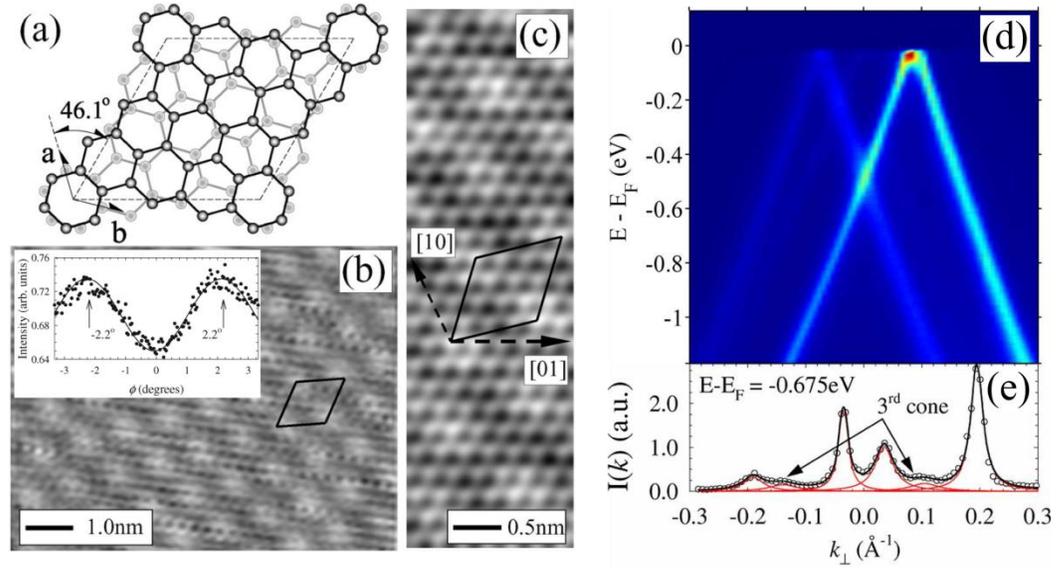

**Figure 4**

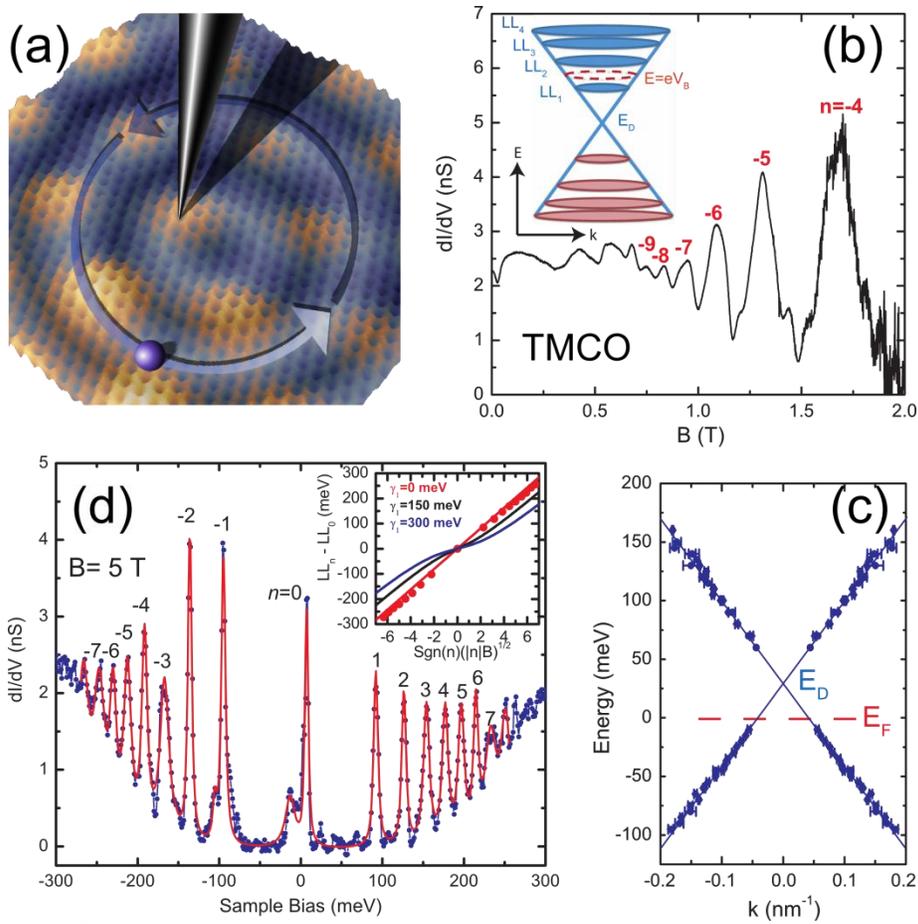

**Figure 5**

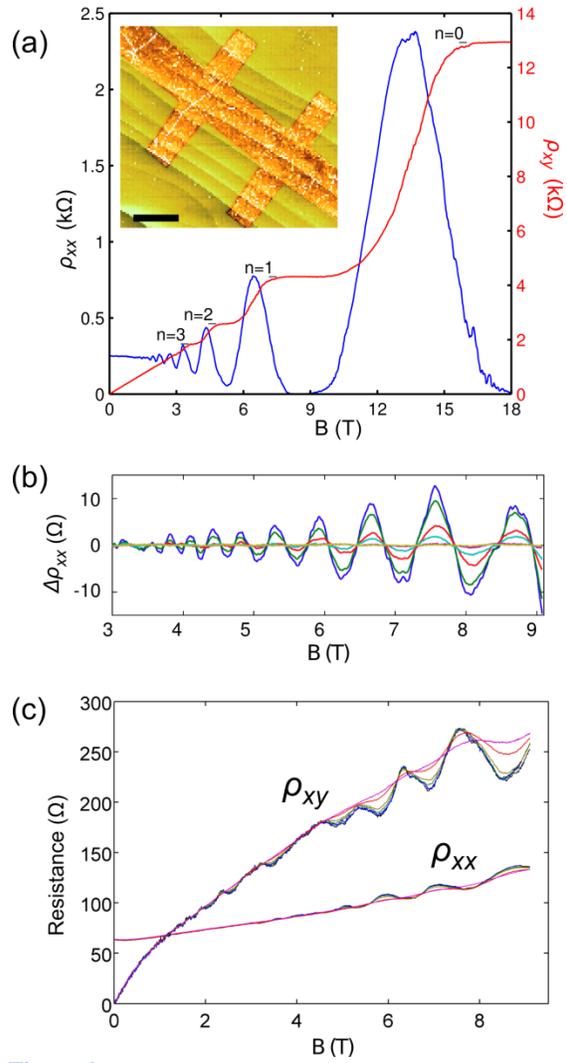

**Figure 6**

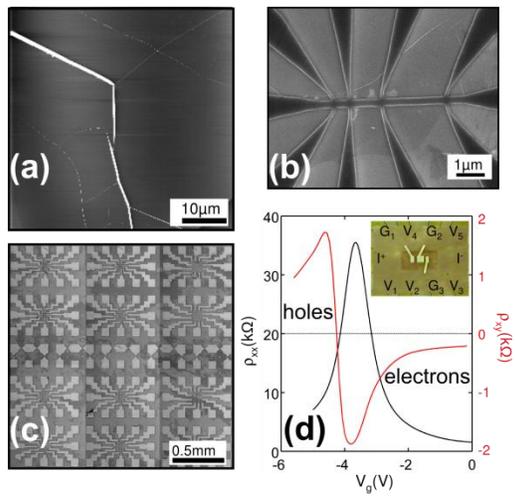

**Figure 7**

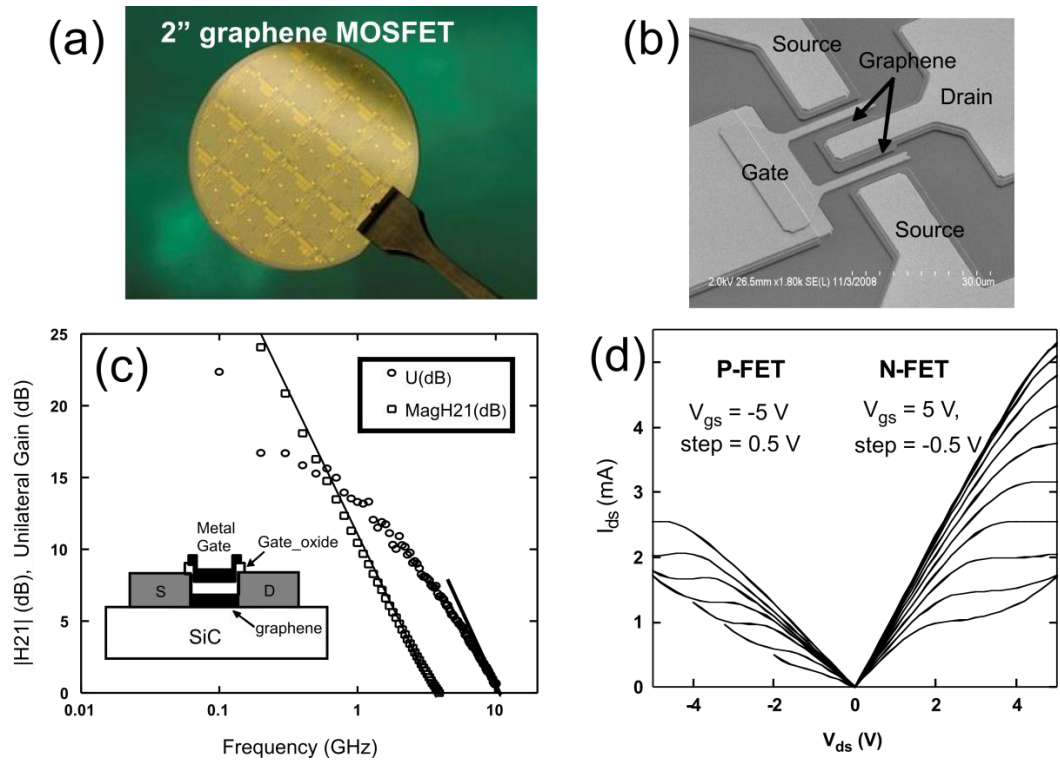

**Figure 8**

**Author Biographies**

Phillip N. First can be contacted at the School of Physics, Georgia Institute of Technology, Atlanta, GA  30332-0430, and [first@gatech.edu](mailto:first@gatech.edu).

Professor First is a faculty member in the School of Physics at Georgia Tech.  He earned a B.S. degree in physics from the University of Wisconsin and a Ph.D. from the University of Illinois in 1988.  His research has been in the physics of surfaces, interfaces, and nanostructures, including the recent development of epitaxial graphene as an electronic material.  First also was instrumental in establishing the Magnetic Interfaces and Nanostructures Division of the AVS, serving as its founding Chair.

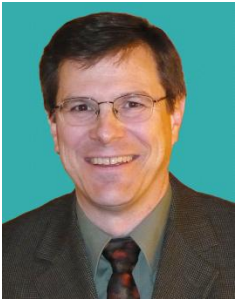

Walt A. de Heer can be contacted at the School of Physics, Georgia Institute of Technology, Atlanta, GA  30332-0430, and walt.deheer@physics.gatech.edu.

De Heer received his Ph.D. in Physics from UC-Berkeley in 1985 and continued at Berkeley as a Postdoctoral Fellow before joining the EPFL in Lausanne, Switzerland.  In 1996 he moved to Georgia Tech, where he is now a Regent's Professor of Physics.  De Heer is a Fellow of the APS, recognized for his pioneering work in metal clusters and nanotubes.  His development of nanopatterned epitaxial graphene for nanoelectronics was named one of MIT Technology Review's "Top 10 Technology Breakthroughs" in 2008.

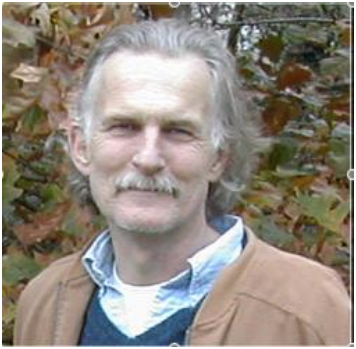

Thomas Seyller can be contacted at Lehrstuhl für Technische Physik, Universität Erlangen-Nürnberg, Erwin-Rommel-Str. 1, 91058 Erlangen, Germany, and Thomas.Seyller@physik.uni-erlangen.de.

Seyller earned his Diploma in Physics (1993) and Ph.D. in Physical Chemistry (1996) from the University of Erlangen-Nürnberg. Following postdoctoral work at Pennsylvania State University, he returned to the Chair of Technical Physics, at the University of Erlangen-Nürnberg as a scientific research associate. In 2006, Seyller completed his Habilitation in Physics. He is now a senior research associate at the Chair of Technical Physics at Erlangen-Nürnberg. His research interests lie in the growth and characterization of electronic materials, particularly SiC and epitaxial graphene.

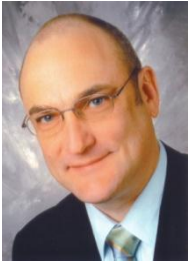

Claire Berger can be reached at CNRS-Institut Néel, 38042 Grenoble Cedex 9, France, and <claire.berger@grenoble.cnrs.fr>

Dr. Berger received the Ph.D. degree in physics from the University Joseph Fourier, Grenoble, France and was a Postdoctoral Fellow with the Center for Atomic Studies, France. She joined the French National Center for Scientific Research (CNRS) in Grenoble, where her main interest was electronic properties of quasicrystals. She presently is also a visiting research scientist at the Georgia Institute of Technology, where she and coworkers pioneered graphene electronics. Her main focus is the growth and electronic transport properties of epitaxial graphene.

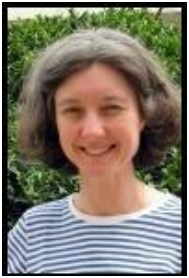

Joseph A. Stroscio can be contacted at the Center for Nanoscale Science and Technology, National Institute of Standards and Technology (NIST), Gaithersburg, MD 20899, USA, and joseph.stroscio@nist.gov

Stroscio received his Ph.D. in Physics from Cornell University. He joined NIST in 1987 after a postdoc at IBM, where he pioneered key developments in scanning tunneling microscopy and spectroscopy. Dr. Stroscio is a Fellow of the American Physical Society and the American Vacuum Society and has received several honors for his research at NIST, including the Department of Commerce Gold Medal Award. His research interests include atomic manipulation, the physics of low-dimensional systems and nanostructures, and nanoscale magnetism.

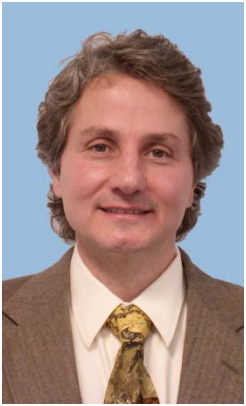

Jeong-Sun Moon can be reached at HRL Laboratories LLC, Malibu, CA 90265, USA, and JMoon@hrl.com.

Moon is a Senior Research Scientist at HRL. In 1995 he received his Ph.D. from Michigan State University, studying quantum devices and digital-signal-processing. Prior to joining HRL, Dr. Moon worked at Sandia National Laboratories.  At HRL, he has worked on optical devices and on emerging materials/devices/RF circuits using GaN, InP, GaSb, SiGe and graphene. Dr. Moon has authored 60 papers and has served as PI on contracts from DARPA, ONR, NRO, JPL and NASA. He holds 9 patents, with 5 pending.

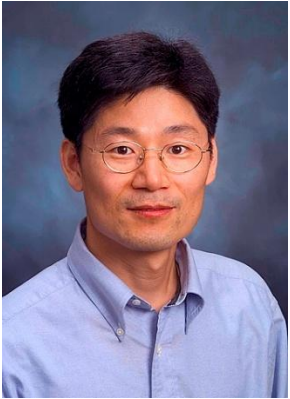